\pdfoutput=1

\documentclass[11pt]{article}

\usepackage{emnlp2021}
\usepackage{times}
\usepackage{latexsym}
\usepackage[T1]{fontenc}
\usepackage[utf8]{inputenc}
\usepackage{microtype}
\usepackage{booktabs}
\usepackage{multirow}
\usepackage{graphicx}
\usepackage{float}
\usepackage{soul}
\usepackage[inline]{enumitem}

\title{Combining~Lexical~and~Dense~Retrieval for~Computationally~Efficient~Multi-hop~Question~Answering}

\author{
        Georgios Sidiropoulos$^{1}$,
        Nikos Voskarides$^{2} $\thanks{\hspace{2mm}Research conducted when the author was at the University of Amsterdam.} ,
        Svitlana Vakulenko$^{1}$,
        Evangelos Kanoulas$^{1}$ \\
        $^1$ University of Amsterdam, Amsterdam, The Netherlands \\
        $^2$ Amazon, Barcelona, Spain
    {\tt}\\
        {\tt g.sidiropoulos@uva.nl, nvvoskar@amazon.com, s.vakulenko@uva.nl,} \\
        {\tt e.kanoulas@uva.nl}
        }

\begin{document}
\maketitle
\begin{abstract}
In simple open-domain question answering (QA), dense retrieval has become one of the standard approaches for retrieving the relevant passages to infer an answer. Recently, dense retrieval also achieved state-of-the-art results in multi-hop QA, where aggregating information from multiple pieces of information and reasoning over them is required. Despite their success, dense retrieval methods are computationally intensive, requiring multiple GPUs to train. In this work, we introduce a hybrid (lexical and dense) retrieval approach that is highly competitive with the state-of-the-art dense retrieval models, while requiring substantially less computational resources.
Additionally, we provide an in-depth evaluation of dense retrieval methods on limited computational resource settings, something that is missing from the current literature.
\end{abstract}

\section{Introduction}
Multi-hop QA requires retrieval and reasoning over multiple pieces of information~\cite{yang2018hotpotqa}. For instance, consider the multi-hop question: ``Where is the multinational company founded by Robert Smith headquartered?''. To answer this question we first need to retrieve the passage about Robert Smith in order to find the name of the company he founded (General Mills), and subsequently retrieve the passage about General Mills, which contains the answer the question (Golden Valley, Minnesota). Even though multi-hop QA requires multiple retrieval hops, it is fundamentally different from session search~\cite{yang2015query,levine2017extended} and conversational search~\cite{dalton2020trec,voskarides2020query,vakulenko2021question}, since in multi-hop QA the information need of the user is expressed in a single question, thus not requiring multiple turns of interaction.

QA systems typically consist of (i) a retriever that identifies the passage/document in the underlying collection that contains the answer to the user's question, and (ii) a reader that extracts or generates the answer from the identified passage~\cite{Chen2017ReadingWT}. Given that often the answer cannot be found in the top-ranked passage, inference follows a standard beam-search procedure, where top-$k$ passages are retrieved and the reader scores are computed for all $k$ passages~\cite{lee2019latent}. 
However, readers are very sensitive to noise in the top-$k$ passages, thus making the performance of the retriever critical for the performance of QA systems~\cite{yang2019end}.
This is further amplified in multi-hop QA, where multiple retrieval hops are performed; potential retrieval errors get propagated across hops and thus severely harm QA performance.

The majority of current approaches to multi-hop QA use either traditional IR methods (TF-IDF, BM25) \cite{qi2019answering} or graph-based methods for the retriever \cite{nie2019revealing, asai2019learning}. However, those approaches have serious limitations. The former approaches require high lexical overlap between questions and relevant passages, while the latter rely on an interlinked underlying corpus, which is not always the case. 
Recently, \citet{xiong2020answering} introduced a dense multi-hop  passage retrieval model that constructs a new query representation based on the question and previously retrieved passages and subsequently uses the new representation to retrieve the next set of relevant passages.
This model achieved state-of-the-art results while not relying on an interlinked underlying  corpus.

Even though dense retrieval models achieve state-of-the-art results on multi-hop QA, they are computationally intensive, requiring multiple GPUs to train. Existing work only reports results for the cases where such resources are available; therefore providing no answer on how feasible is to train such models on a low resource setting.
In this paper, we focus on developing an efficient retriever for multi-hop QA that can be trained effectively in a low computational resource setting. We aim to answer the following research questions:
\begin{enumerate}[label=\textbf{RQ\arabic*},leftmargin=*]
\item How does the performance of dense retrieval compare to lexical and hybrid approaches?
\item How does the performance degrade in the low computational resource settings?
\end{enumerate} 

Our main contributions are the following: 
(i) we propose a hybrid (lexical and dense) retrieval model which is competitive against its fully dense competitors while requiring eight times less computational power and (ii) we perform a thorough analysis on the performance of dense passage retrieval models on the task of multi-hop QA.\footnote{Our trained models and data are available at \url{https://github.com/GSidiropoulos/hybrid_retrieval_for_efficient_qa}.}

\section{Task Description}
\label{sec:task-desc}
Let $p \in \mathcal{C}$ denote a passage within a passage collection $\mathcal{C}$, and $q$ a multi-hop question.
Given $q$ and $\mathcal{C}$ the task is to retrieve a set of relevant passages $\mathcal{P} = \{p_1,p_2, \dots, p_n\}$, where $p_i \in \mathcal{C}$.
In the multi-hop scenario we consider here, not all relevant passages can be retrieved using the input question $q$ alone. 
This is due to the fact that there is a low lexical overlap or semantic relationship between question $q$ and one or more of the relevant passages in $\mathcal{P}$. In this case, information from one of the relevant passages $p_i$ is needed to retrieve another relevant passage $p_j$, where $p_j$ may be lexically/semantically different from question $q$.

\section{Experimental Setup}
In this section, we describe the dataset used in our experiments, the metrics we use to answer our research questions and the models we compare against.

\subsection{Dataset}
For our experiments, we focus on the HotpotQA dataset and particularly the full-wiki setting~\cite{yang2018hotpotqa}. HotpotQA is a large-scale 2-hop QA dataset where the answers to questions must be found in the context of the entire Wikipedia. Questions in HotpotQA fall into one of the following categories: bridge or comparison. In \emph{bridge} questions, the bridge entity that connects the two relevant passages is missing; e.g., `When did the show that Skeet Ulrich is currently starring in premiere?'', where the bridge entity ``Riverdale (2017 TV series)'' is missing. In \emph{comparison} questions the main two entities (of the two relevant passages) are both mentioned and compared; e.g., ``Which has smaller flowers, Campsis or Kalmiopsis?''. 

\subsection{Metrics}
Following previous work \cite{yang2018hotpotqa, xiong2020answering}, we report passage Exact Match (EM) to measure the overall retrieval performance. Exact Match is a metric that evaluates whether both of the ground-truth passages for each question are included in the retrieved passages (then EM=1 otherwise EM=0). Note that metrics such as EM and F1 w.r.t question's answer (Ans) and supporting facts on sentence-level (Sup) do not fit in our experimental setup since we focus on the retrieval part of the pipeline and not on the reading.  

\subsection{Models}
\label{sec:models}
In this section, we describe the models we experiment with.
    
\subsubsection{Single-hop models}
\label{sec:single-hop}
Given a question, single-hop models retrieve a ranked list of passages. Thus, they are not aware of the multi-hop nature of the task. 

\textbf{BM25} is a standard lexical retrieval model. We use the default Anserini parameters~\cite{yang2017anserini}.

\textbf{Rerank} is a standard two-stage retrieval model that first retrieves passages with BM25 and then uses BERT to rerank the top-k passages~\cite{DBLP:journals/corr/abs-1901-04085}. The BERT (base) classifier was trained with a point-wise loss~\cite{DBLP:journals/corr/abs-1901-04085}. It was fine-tuned on the train split of HotpotQA for 2 epochs.
Training took 5 hours with batch size of 8 using a single 12GB GPU.
We experimented with k=100 and k=1000, and found that k=100 results in a better reranking performance at the top positions.

\textbf{DPR} is a dense passage retrieval model for simple questions \cite{karpukhin2020dense}. Given a question $q$, a relevant passage $p^+$ and a set of irrelevant passages $\{p_1^-, p_2^-, \dots, p_m^-\}$,  the model learns to rank $p^+$ higher via the optimization of the negative log likelihood of the relevant passage. To train DPR on HotpotQA, a multi-hop QA dataset, we follow the procedure described in  \cite{xiong2020answering}. This model was trained for 25 epochs ($\sim 2$ days) on a single 12GB GPU, using a RoBERTa-based encoder. 

\subsubsection{Multi-hop models} These models are aware of the multi-hop nature of the task. They recursively retrieve new information at each hop by conditioning the question on information retrieved on previous hops~\cite{xiong2020answering}. In practice, at each hop $t$ the question $q$ and the passage retrieved in the previous hop $p_{t-1}$ gets encoded as the new query $q_t = h(q, p_{t-1})$, where $h(\cdot)$ the question encoder, to retrieve the next relevant passage; when $t=1$ then we have just the question.
Differently from single-hop models, at inference time, given a question, beam search is used to obtain the top-k passage pair candidates. The candidates to beam search at each hop are generated by a similarity function using the query representation at hop $t$, and the beams are scored by the sum of the individual similarity scores.

\label{subsubsection:multi-hop_models}
\textbf{MDR} is a state-of-the-art dense retriever for multi-hop questions~\cite{xiong2020answering}. It extends DPR in an iterative fashion by encoding the question and passages retrieved in previous hops as the new query to retrieve the next relevant passages. This model was trained for 25 epochs ($\sim3$ days) on a single 12GB GPU, using a RoBERTa-based encoder, without the memory bank mechanism \cite{wu2018unsupervised}. The memory bank mechanism is dropped since it is very expensive to compute and its contribution to retrieval performance is limited.
    
\textbf{MDR (full)} is MDR with the additional memory bank mechanism, trained for 50 epochs on 8$\times$32GB GPUs by ~\citet{xiong2020answering}.

\textbf{Rerank + DPR$_2$} is a hybrid model we propose in this paper. Specifically, for the first hop we rely on the BERT-based re-ranking model described in Section~\ref{sec:single-hop} (Rerank), while for the second hop we train a DPR only on second hop questions (DPR$_2$). To train the latter, we build a variation of HotpotQA where the question gets concatenated with the ground truth passage of the first hop, and the second hop ground truth passage is the only relevant passage to be retrieved. DPR$_2$ was trained for 25 epochs ($\sim1$ day) on a single 12GB GPU, using a RoBERTa-based encoder.

\section{Results}
In this section, we present our experimental results that answer our research questions.
\subsection{Overall performance}
\label{sec:overall}

\begin{table}[t]
\begin{tabular}{@{}lccc@{}}
\toprule
Model & EM@2 & EM@10 & EM@20 \\ \midrule
BM25 & 0.127 & 0.320 & 0.395 \\
Rerank & 0.314 & 0.476 & 0.517 \\
DPR & 0.116 & 0.275 & 0.336 \\ \midrule
MDR & 0.440 & 0.581 & 0.619 \\
Rerank+DPR$_2$ & 0.599 & 0.732 & 0.762 \\ 
MDR (full) & \bf 0.677 &  \bf 0.772 & \bf 0.793\\ \bottomrule
\end{tabular}
\caption{Overall retrieval performance. In the first group we show single-hop models, while in the second group we show multi-hop models.}
\label{tab:overall}
\end{table}

Here we aim to answer \textbf{RQ1} by comparing the retrieval performance of the models we consider.
In Table~\ref{tab:overall}, we observe that the single-hop models perform much worse than the multi-hop models.
This is expected since single-hop models are not aware of the multi-hop nature of the task.

As for the multi-hop models, we observe that MDR (full) achieves the best performance at the higher positions in the ranking.
It is important to underline here that MDR (full) uses considerably more resources than Rerank+DPR$_2$ and MDR. The last two use relatively limited computational resources and a comparison between them is more fair (see Section~\ref{sec:models}).\footnote{Even though the memory bank mechanism is omitted from MDR, the comparison of Rerank+DPR$_2$ and MDR remains fair since this particular mechanism can also be potentially applied to Rerank+DPR$_2$ (in the DPR part).}
We observe that our Rerank+DPR$_2$ outperforms MDR on all metrics while is also competitive against MDR (full), especially w.r.t EM@10 and EM@20.
This is due to the fact that often questions and their relevant passages are not only semantically related, but also have high lexical overlap.  This is also highlighted by \citet{karpukhin2020dense}, who reported that dense retrieval has performance issues when the question has high lexical overlap with the passages.

\begin{table*}[t]
\resizebox{\textwidth}{!}{%
\centering
\begin{tabular}{llccccrrr}
\toprule
Model & Encoder & \# GPU & \# Epochs & Batch size & \# Gradient acc. steps & \multicolumn{1}{l}{EM@2} & \multicolumn{1}{l}{EM@10} & \multicolumn{1}{l}{EM@20} \\ \midrule
MDR (full) & RoBERTa & 8  $\times$ 32GB & 50 & 150 & 1 & 0.677 & 0.772 & 0.793 \\ \midrule
MDR & RoBERTa & 8  $\times$ 32GB & 50 & 150 & 1 & 0.637 & 0.742 & 0.772 \\
 & RoBERTa & 4 $\times$ 24GB & 50 & 28 & 1 & 0.606 & 0.711 & 0.735 \\
 & RoBERTa & 4 $\times$ 24GB & 25 & 28 & 1 & 0.550 & 0.668 & 0.698 \\
 & RoBERTa & 4 $\times$ 24GB & 20 & 28 & 1 & 0.537 & 0.659 & 0.687 \\
 & RoBERTa & 1 $\times$ 12GB & 25 & 4 & 32 & 0.440 & 0.581 & 0.619 \\
 & BERT & 1 $\times$ 12GB & 25 & 4 & 32 & 0.421 & 0.560 & 0.597 \\ \midrule
DPR & RoBERTa & 8  $\times$ 32GB & 50 & 256 & 1 & 0.252 & 0.454 & 0.521 \\
 & RoBERTa & 4 $\times$ 24GB & 25 & 128 & 1 & 0.223 & 0.427 & 0.487 \\
 & RoBERTa & 1 $\times$ 12GB & 25 & 8 & 32 & 0.116 & 0.275 & 0.336 \\ \bottomrule
\end{tabular}
}
\caption{Analysis of how computational resources affect the performance of MDR and DPR. The MDR(full) configuration is provided by \cite{xiong2020answering}. Different beam size can slightly change the results from what was originally reported. MDR and DPR, both trained trained on 8 GPUs, are not available and therefore we report the results as they were reported in \cite{xiong2020answering}.}
\label{tab:resources}
\end{table*}

\subsection{Performance for limited resources}
\label{sec:limited-resources}
In this section, we answer \textbf{RQ2~} by comparing the retrieval performance of MDR and DPR as provided in \cite{xiong2020answering}, against the same models trained with limited computational resources.

In Table ~\ref{tab:resources} we see that performance drops significantly as we limit resources for both DPR and MDR. This is a result of the training scheme that is used in \cite{karpukhin2020dense} and \cite{xiong2020answering}. More specifically, DPR and MDR rely on using in-batch negatives both for decreasing the training time (positive passages of a question are reused as negative passages for the rest of the questions in the batch, instead of having to sample new ones beforehand), and for improving accuracy (bigger batch size will produce more in-batch negatives, thus increasing the number of training samples).
When we have limited resources, training time gets significantly longer (since we use fewer GPUs), and therefore we have to compromise for fewer training epochs while the batch size is restricted by the GPU memory size. 
For instance, training for 50 epochs takes $\sim$1 day on 8$\times$32GB GPUs, while it takes $\sim$6 days on a single 12GB GPU. 

In addition, when comparing MDR trained on  4$\times$24GB GPUs against MDR trained on a single 12GB GPU, for 25 epochs each, we observe that even though we can simulate bigger batch sizes by using gradient accumulation,\footnote{Gradient accumulation is a mechanism that accumulates the gradients and the losses over consecutive batches for a specific number of steps (without updating the parameters), and then updates the parameters based on the cumulative gradient~\cite{mnih2016asynchronous}.} we do not observe an increase in performance. This is a consequence of the fact that the number of in-batch negatives is limited by the real batch size. Note that we also observe a similar trend for DPR.

In summary, computational resources are of vital importance for multi-hop dense retrieval models. Hence, in the case where only limited resources are available, following a hybrid (lexical and dense) approach such as our proposed Rerank+DPR$_2$ seems to be a good choice. As we showed in Tables \ref{tab:overall} and \ref{tab:resources}, Rerank+DPR$_2$ (trained on a single GPU) performs similarly to MDR trained on 4 GPUs and is competitive against MDR trained on 8 GPUs.

\begin{table*}[!t]
\resizebox{\textwidth}{!}{%
\begin{tabular}{@{}ll@{}}
\toprule
& ${q}$ \textbf{[bridge]: Who was the defense counsel of a German woman who underwent Catholic exorcism rites during the year before her death?} \\ \toprule
\multicolumn{1}{l|}{\multirow{2}{*}{Rerank+DPR$_2$}} & \begin{tabular}[c]{@{}l@{}} \textbf{Anneliese Michel}: Anneliese Michel {]} (21 September 1952 – 1 July 1976) was a \underline{German woman who underwent Catholic exorcism rites during the year before her death}.\\ Later investigation determined that she was malnourished and dehydrated\dots \end{tabular} \\
\multicolumn{1}{l|}{} & \begin{tabular}[c]{@{}l@{}}\textbf{Erich Schmidt-Leichner}: \hl{Erich Schmidt-Leichner} (14 October 1910 – 17 March 1983) was a German lawyer who made a name as a distinguished defense counsel at the\\ Nuremberg Trials (1945 - 1946). In 1978, he was a defense counsel in the "Klingenberg Case" (Anneliese Michel)\dots \end{tabular} \\ \midrule
\multicolumn{1}{l|}{\multirow{2}{*}{MDR}} & \textbf{Maria Pauer}: Maria Pauer (October 1734/36 – 1750 in Salzburg), was an alleged Austrian witch.  She was the last person to be executed for witchcraft in Austria. \\
\multicolumn{1}{l|}{} & \textbf{Franz Burkard (died 1539)}: Franz Burkard (died 1539) was a canon lawyer in Ingoldstadt who opposed Lutheranism, particularly in the trial of Andreas Seehofer. \\ \bottomrule
\toprule
& ${q}$ \textbf{[bridge]: The astronomer who formulated the theory of stellar nucleosynthesis co-authored what landmark paper?} \\ \toprule
\multicolumn{1}{l|}{\multirow{2}{*}{Rerank+DPR$_2$}} & \textbf{Fred Hoyle}: Sir Fred Hoyle FRS (24 June 1915 – 20 August 2001) was an English \underline{astronomer who formulated the theory of stellar nucleosynthesis} \dots \\
\multicolumn{1}{l|}{} & \begin{tabular}[c]{@{}l@{}}\textbf{B2FH paper}: The \hl{BFH paper}, named after the initials of the authors of the paper, Margaret Burbidge, Geoffrey Burbidge, William Fowler, \\ and Fred Hoyle, is a landmark paper of stellar physics published in "Reviews of Modern Physics" in 1957 \dots \end{tabular} \\ \midrule
\multicolumn{1}{l|}{\multirow{2}{*}{MDR}} & \textbf{Sequence hypothesis}: The sequence hypothesis was first formally proposed in the review "On Protein Synthesis" by Francis Crick in 1958 \dots \\
\multicolumn{1}{l|}{} & \begin{tabular}[c]{@{}l@{}}\textbf{Francis Crick}: Francis Harry Compton Crick (8 June 1916 – 28 July 2004) was a British molecular biologist, biophysicist, and neuroscientist,\\  most noted for being a co-discoverer of the structure of the DNA molecule in \dots \end{tabular} \\ \bottomrule
\end{tabular}%
}
\caption{Two example bridge questions for which Rerank+DPR$_2$ retrieves both relevant passages at the top positions while MDR fails to do so. Lexical overlap is indicated with underlined text while the answer to the question is highlighted.}
\label{tab:hybrid_wins}
\end{table*}

\begin{table*}[!t]
\resizebox{\textwidth}{!}{%
\begin{tabular}{@{}ll@{}}
\toprule
 & ${q}$ \textbf{[bridge]: What Golden Globe Award actor starred in the film Little Fugitive?} \\ \toprule
\multicolumn{1}{l|}{\multirow{2}{*}{Rerank+DPR$_2$}} & \begin{tabular}[c]{@{}l@{}} \textbf{Ali MacGraw:} Elizabeth Alice "Ali" MacGraw (born April 1, 1939) is an American actress, model, author, and animal rights activist.  She first gained\\ attention with her role in the 1969 film "Goodbye, Columbus", for which she won the Golden Globe Award for Most Promising Newcomer.\\ She reached international fame in 1970's "Love Story", for which she was nominated for an Academy Award for Best Actress and won the\\ Golden Globe Award for Best Actress in a Motion Picture – Drama \dots \end{tabular} \\
\multicolumn{1}{l|}{} & \begin{tabular}[c]{@{}l@{}} \textbf{Little Fugitive}: Little Fugitive (1953) is an American film written and directed by Raymond Abrashkin (as "Ray Ashley"), Morris Engel and Ruth Orkin,\\ that tells the story of a child alone in Coney Island.\end{tabular} \\ \midrule
\multicolumn{1}{l|}{\multirow{2}{*}{MDR}} & \begin{tabular}[c]{@{}l@{}}\textbf{Little Fugitive (2006 film)}: Little Fugitive is a 2006 remake of the film of the same name.  It was directed by Joanna Lipper and produced by Nicholas Paleologos.\\ The film is set in present day Brooklyn and tells the story of 11-year-old Lenny (Nicolas Martí Salgado) who must take care of his 7-year-old brother, Joey (David Castro),\\ while their father (Peter Dinklage) is in jail and their mother work long hours a nursing home \dots \end{tabular} \\
\multicolumn{1}{l|}{} & \begin{tabular}[c]{@{}l@{}}\textbf{Peter Dinklage}: \hl{Peter Hayden Dinklage} ( , born June 11, 1969) is an American actor and film producer.  He has received numerous accolades, including a\\ Golden Globe Award and two Primetime Emmy Awards.\end{tabular} \\ \bottomrule
\toprule
 & ${q}$ \textbf{[bridge]:Which American professional poker player also starred in the 2015 movie "Extraction"?} \\ \toprule
\multicolumn{1}{l|}{\multirow{2}{*}{Rerank+DPR$_2$}} & \textbf{Allen Cunningham}: Allen Cunningham (born March 28, 1977) is an American professional poker player who has won five World Series of Poker bracelets. \\
\multicolumn{1}{l|}{} & \begin{tabular}[c]{@{}l@{}}\textbf{Extraction (film)}: Extraction is a 2015 American action-thriller film directed by Steven C. Miller and written by Umair Aleem.  The film stars Kellan Lutz,\\ Bruce Willis, Gina Carano, D. B. Sweeney, Dan Bilzerian and Steve Coulter \dots \end{tabular} \\ \midrule
\multicolumn{1}{l|}{\multirow{2}{*}{MDR}} & \begin{tabular}[c]{@{}l@{}} \textbf{Extraction (film)}: Extraction is a 2015 American action-thriller film directed by Steven C. Miller and written by Umair Aleem.  The film stars Kellan Lutz,\\ Bruce Willis, Gina Carano, D. B. Sweeney, Dan Bilzerian and Steve Coulter \dots \end{tabular} \\
\multicolumn{1}{l|}{} & \textbf{Dan Bilzerian}: \hl{Dan Brandon Bilzerian} (born December 7, 1980) is an American professional poker player. \\ \bottomrule
\end{tabular}%
}
\caption{Two example bridge questions for which MDR retrieves both relevant passages at the top positions while Rerank+DPR$_2$ fails to do so. The answer to the question is highlighted.}

\label{tab:mdr_wins}
\end{table*}

\subsection{Error Analysis}
\label{sec:analysis}
We perform qualitative analysis to gain further insights into where the models succeed or fail. More specifically, we compare specific cases where our hybrid model (Rerank+DPR$_2$) retrieves both relevant passages successfully while MDR fails and vice versa. For our analysis we focus on bridge questions since comparison questions are more straightforward to retrieve. 

Table \ref{tab:hybrid_wins} shows two typical examples of questions for which Rerank+DPR$_2$ retrieves both relevant passages at the top positions while MDR fails to do so. When there is a high lexical overlap between the question and a relevant passage, our hybrid model can capture this exact n-gram match and improve the performance. In contrast, fully dense models seem incapable of capturing this. In particular, this lexical overlap can be between the question and both relevant passages for the case of comparison questions, or between the question and the first relevant passage for bridge questions. 

In bridge questions if the lexical overlap is between the question and the second passage then our hybrid model favors passages in which this phrase appears, and therefore it retrieves an irrelevant first passage; leading to an irrelevant second passage as well. MDR on the other hand manages to retrieve both relevant passages at the top positions. Those are the cases where the lexical overlap is used in the given question in order to disambiguate the final answer. Examples can be found in Table \ref{tab:mdr_wins}.

In the first example, ``Golden Globe Award'' is used in the given question in order to disambiguate the final answer, since in the film ``Little Fugitive'' there is more than one actor involved.
Therefore, ``Golden Globe Award'' must be used to assist the retrieval of the second passage.
Since Rerank+DPR$_2$ builds on top of BM25, it favors passages in which this phrase appears, and therefore it retrieves an irrelevant first passage leading to an irrelevant second passage as well. On the other hand, MDR manages to retrieve both relevant passages at the top positions.
Similarly, for the second example, ``American professional poker player'' is used to specify the actor that starred in the ``Extraction'' movie, hence supporting the retrieval of the second relevant passage.

\section{Conclusion}
\label{sec:conclusion}
In this work, we provided insights on the performance of state-of-the-art dense retrieval for multi-hop questions.
We showed that Rerank+DPR$_2$ (our hybrid model) outperforms MDR (the state-of-the-art multi-hop dense retrieval model) in the low resource setting, and it is competitive with MDR in the setting where MDR uses considerably more computational resources. 
Finally, we highlighted that fully dense retrieval models get harmed when using limited computational resources.
For future work, we plan to build on our insights to improve the performance of multi-hop models by combining the strengths of lexical and dense retrieval. Also, we aim to develop less computationally expensive multi-hop retrieval models.

\section*{Acknowledgements}
This research was supported by
the NWO Innovational Research Incentives Scheme Vidi (016.Vidi.189.039),
the NWO Smart Culture - Big Data / Digital Humanities (314-99-301),
the H2020-EU.3.4. - SOCIETAL CHALLENGES - Smart, Green And Integrated Transport (814961).
All content represents the opinion of the authors, which is not necessarily shared or endorsed by their respective employers and/or sponsors.

\bibliography{anthology,custom}
\bibliographystyle{acl_natbib}

\end{document}